\documentclass[11pt]{article}

\usepackage[USenglish]{babel}
\usepackage[utf8]{inputenc}
\usepackage[T1]{fontenc}
\usepackage{lmodern}
\usepackage{hyperref}
\usepackage{microtype}
\usepackage[margin=1in]{geometry}
\usepackage{amsmath,amssymb,amsthm}
\usepackage{graphicx}
\usepackage[authoryear,round]{natbib}
\usepackage{longtable}
\usepackage{booktabs}

\theoremstyle{definition}

\theoremstyle{remark}

\newcommand{\bu}{\mathbf{u}}
\newcommand{\bz}{\mathbf{z}}
\newcommand{\bmu}{\boldsymbol{\mu}}

\title{Clustering professional football players using Gaussian mixture models:\\
A case study of the English Premier League 2024/25}

\author{%
Harry O'Donnell\\
\small Department of Computer Science, University of Bath, Bath, UK\\
\small \texttt{hod25@bath.ac.uk}
\and
Antony M. Overstall\\
\small School of Mathematical Sciences, University of Southampton, Southampton, UK\\
\small \texttt{A.M.Overstall@soton.ac.uk}\\
\small ORCID: 0000-0003-0638-8635
}
\date{}

\begin{document}
\maketitle

\begin{abstract}
The increasing availability of detailed football performance data has created new opportunities for identifying player archetypes and informing recruitment and tactical decision-making. Most existing clustering approaches assign each player to a single group, despite the reality that many modern footballers perform hybrid roles spanning multiple stylistic categories. Performance data for 314 outfield players from the 2024/25 English Premier League season are analysed using a Gaussian mixture model (GMM) applied to principal component representations of player statistics. Unlike hard-assignment methods such as k-means, the GMM provides probabilistic cluster memberships, allowing uncertainty and role hybridity to be quantified. Seven player archetypes are identified in the full dataset. Posterior membership probabilities reveal a subset of players whose profiles span multiple clusters, providing an objective measure of tactical flexibility and stylistic ambiguity. Further analyses within defender, midfielder, and forward groups recover position-specific sub-archetypes and highlight exceptional outliers. The results demonstrate that model-based clustering provides a flexible and interpretable framework for identifying latent structure in high-dimensional performance data while explicitly quantifying uncertainty in cluster membership. More broadly, the approach is readily transferable to other sports and domains in which individuals exhibit overlapping characteristics and cannot be adequately represented by hard cluster assignments alone.
\end{abstract}


\section{Introduction} 

\subsection{Overview}

The growing availability of detailed player performance data has changed how football clubs approach recruitment and tactical planning. Metrics capturing everything from progressive carries to defensive actions are now collected across all major leagues, yet summarising these profiles into interpretable player archetypes remains an open problem.

Clustering methods offer one approach. Applied to football data, they have identified broad positional groupings and stylistic patterns that map onto recognisable tactical roles \citep{rein2016, decroos2020, akhanli2023}. A common limitation of existing approaches, however, is that each player receives a single cluster label, a constraint that misrepresents the reality of modern football, where many players carry out responsibilities that cut across multiple stylistic categories.

This paper addresses this limitation by applying a Gaussian Mixture Model (GMM; \citealt{dempster1977}) to Premier League outfield player performance data from the 2024--25 season. Unlike k-means \citep{hartiganwong1979}, which assigns each player to exactly one cluster, a GMM produces soft cluster assignments by estimating the probability that each player belongs to each cluster. This makes it possible to identify hybrid roles and measure how mixed they are. The analysis can also be applied separately to pre-defined positional groupings (i.e. defenders, midfielders and forwards) to identify sub-archetypes within each group.

\subsection{Data} \label{sec:data}

As a case study, player performance data were collected from the 2024--25 Premier League season using the FBref database \citep{fbref2026}. Statistics were extracted from eight tables: standard statistics, shooting, passing, pass types, goal and shot creation, defensive actions, possession, and miscellaneous metrics. Goalkeepers were excluded from the analysis due to their fundamentally different performance profile. Players with fewer than 800 minutes (approximately nine full matches) played were removed so that each profile is based on a reasonable amount of playing time. After filtering, the dataset comprised of $n=314$ outfield player records. Of these, and for the within-position analysis, FBref identified $n_D = 142$ defenders, $n_M = 144$ midfielders, and $n_F = 112$ forwards. The reason $n \ne n_D + n_M + n_F$ is because FBref, itself, recognises that some players possess hybrid roles. For example, Marcus Rashford and Kevin De Bruyne appear as both midfielders and forwards. This lends credence to the idea of using clustering to identify hybrid players with an objective, evidence-based approach. 

Variables describing player characteristics or availability instead of playing style (age, year of birth, appearances, starts, minutes played) were excluded from the feature set. The remaining $d=123$ features (see Table SM1 in the Supplementary Material) are season totals, with the exception of rate, percentage and average variables such as pass completion percentage, shots per 90 and average shot distance (see Section~\ref{sec:limitations} for a discussion on the implications of this choice). For within-position clustering, features with little relevance to a position were removed for that position's feature set, e.g. shooting metrics for defenders. Table SM1 in the Supplementary Material shows the removed features for each position dataset. The number of features for each position are $d_D = 92$ (for defenders), $d_M = 116$ (for midfielders) and $d_F = 100$ (for forwards). 

%

\section{Methodology} \label{sec:methodology}

\subsection{Principal Component Analysis} \label{sec:pca}

Before applying clustering with GMMs, principal components analysis (PCA; e.g. \citealt{jolliffe2002}) is used to reduce the dimensionality of the feature space. Clustering in the PCA-reduced space diminishes the effect of multicollinearity and mitigates the curse of dimensionality which leads to more reliable distance-based computations, when compared to clustering in the original feature space \citep{beyer1999}. 

Let $X$ denote the $n \times d$ data matrix, where the $ij$th element of $X$ gives the value of the $j$th feature for the $i$th player, for $i=1,\dots,n$ and $j=1,\dots,d$. Standardisation has been applied prior to forming $X$, so that the sample mean and variance of each column of $X$ is zero and one, respectively.

Let $\left\{\bu_1,\dots,\bu_d\right\}$ and $\left\{\lambda_1,\dots,\lambda_d\right\}$ denote the eigen-vectors and -values of the sample correlation matrix $P = \frac{1}{n-1}X^TX$ of $X$. The eigen-vectors and -values are ordered such that $\lambda_1 \ge \lambda_2 \ge \dots \ge \lambda_d$. The PCA-reduced representation of the data $X$ is the $n \times m$ matrix $Z = X U^{(m)}$ where $U$ is the $n \times m$ matrix with columns given by the $m \le d$ leading eigenvectors $\bu_1\dots,\bu_m$. The number of principal components, $m$, is chosen to explain 75\% of the variation, i.e. $m = \min \left\{ \ell : \sum_{j=1}^{\ell} \lambda_j / d \ge 0.75 \right\}$. The choice of 75\% was motivated by stability: with higher-dimensional representations, the GMM produced uninterpretable solutions in which singleton clusters absorb a large proportion of players or clusters contain small numbers of players.

\subsection{Gaussian mixture model}

\subsubsection{Model formulation}

A Gaussian mixture model \citep{dempster1977} is a probabilistic method that assumes the vectors of player features each arise as a mixture of $K$ multivariate Gaussian/normal distributions. Unlike k-means clustering which assigns each player to a single cluster deterministically, a GMM produces a posterior probability distribution over cluster membership of each player, accommodating allocation uncertainty, and allowing cluster overlap.

Let $\bz_i$ be the $i$th row of $Z$, giving the $i$th player's PCA-reduced features, for $i=1,\dots,n$. The vectors $\bz_1,\dots,\bz_n$ are assumed independent, with the probability density function (pdf) of $\bz_i$ given by
\begin{equation}
f(\bz_i) = \sum_{k=1}^K \pi_k \psi (\bz_i ; \bmu_k, \Sigma_k)
\label{eqn:gmm}
\end{equation}
for $i=1,\dots,n$. In (\ref{eqn:gmm}), $\pi_1,\dots,\pi_K$ are the mixture weights with $0 < \pi_k < 1$ and $\sum_{k=1}^K \pi_k = 1$, and $\psi(\cdot; \bmu_k, \Sigma_k)$ is the pdf of the $m$-variate Gaussian distribution with mean vector $\bmu_k$ and variance matrix $\Sigma_k$.

In general, with the only stipulation being that $\Sigma_k$ is positive-definite, there are $p = K - 1 + \frac{1}{2}Km(m+3)$ parameters. The factor $\frac{1}{2}m(m+3)$ refers to the number of unique parameters in $\bmu_k$ and $\Sigma_k$, and $K-1$ to the number of free mixture weights $\pi_1,\dots,\pi_{K-1}$. To simplify, when $n$ is relatively small, a diagonal structure can be imposed on each $\Sigma_k$, leading to $p = K - 1 + 2Km$ parameters.

For a given number of clusters, $K$, and parameters; the posterior probability of the $i$th player belonging to the $k$th cluster is given by Bayes' theorem as
\begin{equation}
\gamma_{ik} = \frac{\pi_k \psi(\bz_i ; \bmu_k, \Sigma_k)}{\sum_{j=1}^K  \pi_j \psi(\bz_i ; \bmu_j, \Sigma_j)},
\label{eqn:postprobs}
\end{equation}
for $i=1,\dots,n$ and $k=1,\dots,K$. 

Whilst the posterior probabilities $\gamma_{i1},\dots,\gamma_{iK}$ provide information about uncertainty in cluster membership for the $i$th player, this information can be summarised by the entropy score \citep[e.g.][page 32]{mackay2003}
\begin{equation}
H_i = - \sum_{k=1}^K \gamma_{ik} \log \gamma_{ik}.
\label{eq:entropy}
\end{equation}
The entropy satisfies $0 \le H_i \le \log K$, with smaller values indicating lower uncertainty and ambiguity in cluster allocation. The higher the entropy, the more likely that the player has a hybrid role. A single summary of cluster membership for the $i$th player can be given by the posterior mode, i.e. $\arg \max_{k=1,\dots,K} \left\{ \gamma_{ik} \right\}$.

\subsubsection{Estimation and cluster selection}

For a given number of clusters, $K$; maximum likelihood is used estimate the $p$ parameters via the expectation-maximisation (EM) algorithm \citep{dempster1977}. 
To mitigate sensitivity to initial parameter values, the EM algorithm is run from 100 random starts. The EM algorithm is implemented by the Python library \texttt{scikit-learn} \citep{pedregosa2011}. The unknown parameter values in (\ref{eqn:postprobs}) can be replaced by their estimates to provide estimates of the posterior probabilities of cluster membership. It follows that the entropy and the posterior mode cluster for each player can also be estimated.

The number of clusters, $K$, is limited from above by how many parameters, $p$, can be estimated by $n$ player observations. Under an unrestricted form for $\Sigma_k$, $K_{\mbox{max}} = \lfloor n/(1+\frac{1}{2}m(m+3)) \rfloor$, and, under a diagonal form,  $K_{\mbox{max}} = \lfloor n/(1+2m) \rfloor$. 

The number of clusters within the set $\left\{ 1, \dots, K_{\mbox{max}}\right\}$ is then selected to minimise the Bayesian information criterion (BIC; \citealt{schwarz1978}). \cite{fraleyraftery2002} advocate using BIC to simultaneously select \emph{both} the number of clusters and the covariance structure (e.g. unrestricted vs diagonal). Due to the small number of players, $n$, in Section~\ref{sec:allgmmcluster}, a more subjective determination is made about the covariance structure to allow a larger number of clusters to be adopted.

\section{Results}

\subsection{All player clustering}

In this section, the results of applying clustering to all $n=314$ players are interpreted.

\subsubsection{Principal component analysis}

The first principal component explains 26.0\% of the total variance in the standardised dataset, while the second explains 24.0\%. Together, the first two components account for approximately 50\% of the total variation in player performance. The third and fourth components explain 7.0\% and 4.9\% respectively. 

The first principal component loads most heavily on chance creation and attacking involvement, including shot-creating actions, touches in the attacking third, key passes, shot-creating actions from live passes and expected assisted goals (xAG). Players with high scores on this axis are typically creative midfielders, inside forwards and attacking wingers, while defenders and deeper midfielders exhibit low scores. This component can therefore be interpreted as an attacking creation dimension.

The second principal component is strongly associated with passing volume, for example passes attempted and completed, total passing distance and live-ball passes, together with touches in the middle third. This suggests that the component reflects overall possession involvement and ball dominance. High scores correspond to players who are heavily involved in build-up play, such as central midfielders and deep-lying playmakers. This component can therefore be interpreted as a ball involvement dimension.

The third principal component loads strongly on aerial duels won and lost, shot distance, non-penalty expected goals per shot and non-penalty expected goals. The pattern points to a distinction based on physical presence and shooting profile, picking out a physical and stylistic attacking dimension. The fourth component is associated with tackles, blocks, fouls committed, and midfield defensive actions, covering defensive intensity and ball-winning activity. Players with high scores on this axis are defensive midfielders, full-backs, and centre-backs.


The significant drop in explained variance after the first two components shows that most of the structural differences between players are captured by just a few key dimensions. Beyond the fifth component, each new component accounts for less than 3\% of the variance. Consequently, a total of $m=8$ principal components explains 75\% of the variability.

\subsubsection{Gaussian mixture model cluster selection} \label{sec:allgmmcluster}

Allowing an unrestricted form for $\Sigma_k$ limits the maximum number of clusters to be only six. Instead, a diagonal structure is imposed on $\Sigma_k$, permitting $K_{\mbox{max}} = 18$. The BIC is minimised by $K=7$ clusters (BIC = 11641.0). The BIC curve (with respect to $K$; not shown) is quite flat in the region of the minimum with the BIC for $K=6$ and $K=8$ being 0.01\% and 0.36\% greater, respectively. Comparing the estimated posterior mode allocation for $K=6$ and $K=7$, the Adjusted Rand Index \citep{hubert1985} is 0.782 indicating broad agreement. Most of the difference comes from a single change: a cluster of the $K = 6$ solution splits into a ball-winning group and a smaller group of set-piece specialists, while the other clusters are largely unchanged. Under the estimated posterior mode allocation, cluster sizes at $K = 7$ range from 19 to 107 players, with no degenerate clusters.

\subsubsection{Cluster interpretation}

\begin{table}[htbp]
\centering
\small
\setlength{\tabcolsep}{3pt}
\begin{tabular}{llll}
\hline
Cluster & Label & Description & Membership \\ \hline
1 & C1 & Reliable defenders & 20 \\
2 & C2 & Ball winners & 51 \\
3 & C3 & Rotational and utility & 107 \\
4 & C4 & Set-piece specialists & 19 \\
5 & C5 & Creative forwards & 50 \\
6 & C6 & Traditional centre-backs & 39 \\
7 & C7 & Penalty area strikers & 28 \\ \hline
\end{tabular}
\caption{Cluster labels, description and membership.}
\label{tab:clusters}
\end{table}

Table~\ref{tab:clusters} shows the cluster labels, description and membership, for each of the $K=7$ clusters. Cluster 1 (\emph{Reliable defenders}) is a group of 20 defenders with a technically reliable, lower-involvement profile, among them Jakub Kiwior, Wesley Fofana, Tosin Adarabioyo, Micky van de Ven and Radu Dragusin. They score highly on medium and total pass completion, take-on success, tackle success and aerial duel win rate, and low on fouls committed, fouls drawn, cards and blocked passes. These defenders are reliable in possession but make relatively few defensive actions.

The 51 players in Cluster 2 (\emph{Ball winners}) are mainly defensive midfielders and full-backs, among them Mois\'{e}s Caicedo, Wilfred Ndidi, Manuel Ugarte, Casemiro, Marc Cucurella and Antonee Robinson. The strongest loadings are on tackles won, tackles in the middle and defensive thirds, and combined tackles and interceptions, while penalty-taking and shooting efficiency are low. The cluster is therefore defined by ball recovery, with limited attacking output.

Cluster 3 (\emph{Rotational and utility}) is the largest and most heterogeneous group at 107 players, spanning lower or more mixed statistical profiles across all positions. Examples are rotational defenders such as Ben White and Riccardo Calafiori, utility midfielders such as James Ward-Prowse and Kobbie Mainoo, and lower-output attackers such as Marcus Rashford and Jordan Ayew. Its z-scores are only modestly positive on shot distance and pass completion and negative on carrying distance, progressive carries, passes received and switches of play, so these players are less involved in possession and are not dominant on the axes that separate the other clusters. Cluster 3 remains large even with $K = 7$ clusters: a sizeable share of Premier League outfield players do not fit any of the six more distinctive archetypes, in line with the continuum of role identity (see Section~\ref{sec:clusteruncert}).

Cluster 4 (\emph{Set-piece specialists}; 19 players) picks out set-piece specialists from central midfield and full-back roles, including Kevin De Bruyne, Bruno Fernandes, Declan Rice, Christian Eriksen, Trent Alexander-Arnold, Andy Robertson and Kieran Trippier. Its highest loadings are on corner kicks taken, both inswinging and outswinging, and on shot-creating and goal-creating actions from dead balls, while long pass completion and shooting efficiency are low. Their creative contribution is concentrated in set-piece delivery, with shots typically taken from lower-quality positions.

Cluster 5 (\emph{Creative forwards}) contains 50 players and consists of wide attackers and creative forwards like Mohamed Salah, Bukayo Saka, Cole Palmer, Phil Foden, Martin \O{}degaard and Gabriel Martinelli. The cluster scores highly on goals and assists above expected values and on goal-creating actions from take-ons and carries into the penalty area, and low on aerial duel success, clearances and shots blocked.

Cluster 6 (\emph{Traditional centre-backs}) is made up almost entirely of 39 traditional centre-backs, including Virgil van Dijk, R\'{u}ben Dias, William Saliba, Ibrahima Konat\'{e} and Gabriel Magalh\~{a}es. It is defined by high clearances, shots blocked and touches in the defensive penalty area and defensive third, and by low shot-creating and goal-creating actions and few take-on attempts, a high-volume, penalty-area defending style.

Cluster 7 (\emph{Penalty area strikers}) is the most distinctive group: 28 high-volume penalty area strikers. Erling Haaland, Alexander Isak, Ollie Watkins, Rasmus H\o{}jlund and Jamie Vardy are all assigned to this cluster. Its players score highly on npxG per shot, shots on target percentage, npxG and offsides, and very low on short, medium and total pass completion, matching players who operate mainly in the penalty area and contribute through direct goal threat as opposed to build-up play.

\begin{figure}[ht]
    \centering
    \includegraphics[width=\columnwidth]{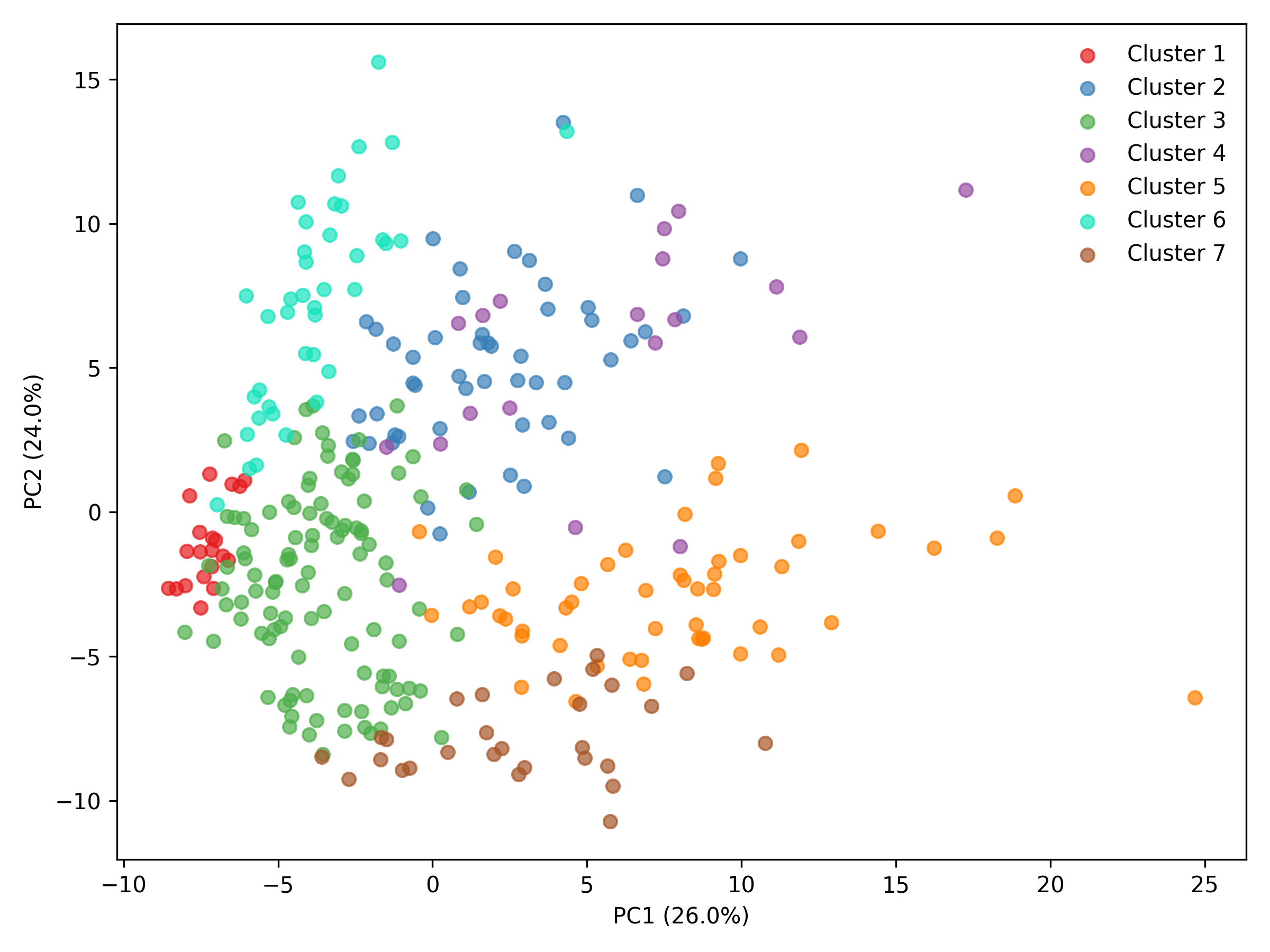}
    \caption{GMM clusters visualised in the space of the first two principal components used for model fitting. Each colour represents a distinct cluster. The remaining six components are not shown but contribute to cluster separation.}
    \label{fig:gmm_clusters}
\end{figure}

The seven clusters broadly align with recognised positional archetypes, but position does not fully determine membership: centre-backs are divided across Clusters 1, 3 and 6 by their style of defensive involvement, and Cluster 4 groups midfielders and full-backs by set-piece responsibility, so playing style also shapes cluster assignment. To support this last statement, Figure~\ref{fig:gmm_clusters} shows a scatterplot of the first two principal components ($\bu_1$ and $\bu_2$) with plotting symbols coloured by cluster membership, where there are clear cluster overlaps.

\subsubsection{Hybrid role identification}
An advantage of probabilistic clustering is the ability to identify players whose profiles span multiple stylistic groups. Players whose membership probabilities exceeded 25\% across more than one cluster were interpreted as hybrid roles, reflecting the tactical flexibility increasingly common in modern football. At this threshold, 24 players were identified as hybrids.

Table~\ref{tab:hybrid_players} presents examples of such players. Mikel Merino is the most ambiguous assignment in the dataset, splitting almost evenly across Cluster 5 (33.9\%; Creative forwards), Cluster 7 (33.6\%; Penalty area strikers) and Cluster 3 (30.4\%; Rotational and utility), reflecting a profile that combines creative attacking output with penalty area presence. Cristian Romero splits almost evenly between Cluster 3 (50.6\%; Rotational and utility) and Cluster 6 (48.7\%; Traditional centre-backs), and Lisandro Mart\'{i}nez shows a similar pattern (66.0\% and 31.6\%), combining a traditional centre-back profile with a more technically involved playing style. Malo Gusto divides between Cluster 3 (52.3\%; Rotational and utility) and Cluster 2 (47.6\%; Ball winners), suggesting that defensively oriented full-back roles sit close to the boundary between these two clusters. Tom\'{a}\v{s} Sou\v{c}ek splits between Cluster 5 (51.9\%; Creative forwards) and Cluster 6 (47.8\%; Traditional centre-backs), combining goal threat from midfield with the aerial and penalty area defending profile of a centre-back. Ryan Manning divides between Cluster 3 (52.2\%; Rotational and utility) and Cluster 4 (44.6\%; Set-piece specialists), reflecting a full-back profile with substantial set-piece responsibility. Dominik Szoboszlai splits between Cluster 2 (61.5\%; Ball winners) and Cluster 5 (38.5\%; Creative forwards), reflecting his dual role as a ball-winning midfielder and a direct attacking contributor. Ashley Young is spread across three clusters, with membership in Cluster 3 (50.5\%; Rotational and utility), Cluster 4 (25.4\%; Set-piece specialists) and Cluster 2 (23.7\%; Ball winners).

\begin{table}[htbp]
\centering
\small
\setlength{\tabcolsep}{3pt}
\begin{tabular}{lccc}
\hline
Player & Primary (\%) & Secondary (\%) & Tertiary (\%)\\
\hline
Mikel Merino & C5 (33.9) & C7 (33.6) & C3 (30.4) \\
Cristian Romero & C3 (50.6) & C6 (48.7) & -- \\
Malo Gusto & C3 (52.3) & C2 (47.6) & -- \\
Tom\'{a}\v{s} Sou\v{c}ek & C5 (51.9) & C6 (47.8) & -- \\
Ryan Manning & C3 (52.2) & C4 (44.6) & -- \\
Dominik Szoboszlai & C2 (61.5) & C5 (38.5) & -- \\
Lisandro Mart\'{i}nez & C3 (66.0) & C6 (31.6) & -- \\
Ashley Young & C3 (50.5) & C4 (25.4) & C2 (23.7) \\
\hline
\end{tabular}
\caption{Example players exhibiting substantial membership across multiple clusters.}
\label{tab:hybrid_players}
\end{table}

Table~SM2 in the Supplementary Material reports all players exhibiting a secondary cluster probability exceeding 10\%.

To assess sensitivity to the choice of threshold, the hybrid player list was evaluated at 20\%, 25\% and 30\%. These yielded 32, 24 and 14 hybrid players respectively. Of the 24 players identified at the 25\% threshold, 14 (58\%) appear at all three thresholds, forming a stable core including Mikel Merino, Cristian Romero, Dominik Szoboszlai, Tom\'{a}\v{s} Sou\v{c}ek and Malo Gusto. The remaining players sit along a gradient of hybrid membership rather than a sharp boundary.

\subsubsection{Cluster uncertainty} \label{sec:clusteruncert}

The maximum posterior probability $\max_k \gamma_{ik}$ was estimated for every player. Across all $n=314$ outfield player records, the median maximum posterior probability was 0.995, indicating that the majority of players were assigned to a primary cluster with strong confidence. The mean maximum posterior probability was 0.939, with a standard deviation of 0.120, so the distribution is concentrated near 1 with a small lower tail. Sixteen players (5.1\%) have a maximum posterior probability below 0.65, and these are the players whose profiles are most ambiguous. The distribution shows no natural discontinuity in this region: the number of players below the threshold rises gradually from 12 (3.8\%) at 0.60 to 16 (5.1\%) at 0.65 and 20 (6.4\%) at 0.70. The threshold of 0.65 was therefore adopted as a pragmatic cut-off, and conclusions drawn from the low-confidence group are not sensitive to its exact value.

Within this group, maximum posterior probabilities ranged from 0.339 (Mikel Merino) to 0.641 (Djed Spence), consistent with the near-even splits identified in Table~\ref{tab:hybrid_players}. The two measures of assignment uncertainty agree closely: 15 of the 16 lowest-confidence players also appear in the hybrid list at the 25\% threshold. The single exception is Facundo Buonanotte (max posterior = 0.626 for Cluster 5 Creative forwards), whose remaining probability is spread across Cluster 2 (20.1\%; Ball winners) and Cluster 3 (16.0\%; Rotational and utility) without either exceeding 25\%. Conversely, nine hybrid players, including Lisandro Mart\'{i}nez (0.660), Ethan Pinnock (0.675), Morgan Gibbs-White (0.721) and Harry Maguire (0.745), were identified by the membership threshold but not by the posterior threshold, illustrating that the two criteria capture overlapping but distinct aspects of allocation ambiguity.


Bilal El Khannouss recorded the highest entropy (see equation \ref{eq:entropy})) of 1.229, with a maximum posterior probability of 0.459 and membership spread across Cluster 2 (45.9\%; Ball winners), Cluster 3 (27.6\%; Rotational and utility), Cluster 5 (18.2\%; Creative forwards) and Cluster 4 (8.3\%; Set piece specialists); no other player is harder to place in a single archetype. Mikel Merino recorded the lowest maximum posterior probability (0.339), with membership split almost evenly between Cluster 5 (33.9\%; Creative forwards), Cluster 7 (33.6\%; Penalty area strikers) and Cluster 3 (30.4\%; Rotational and utility), but only the second-highest entropy (1.178), because his membership falls on three clusters rather than four.

The threshold method shows which clusters a hybrid player sits between, while the maximum posterior probability and entropy summarise uncertainty on a continuous scale for every player.

\subsection{Within-position clustering} \label{sec:within}

In this section, the results of applying clustering to defenders, midfielders and forwards, as identified by FBref, are interpreted.

\subsubsection{Principal components analysis and cluster selection}

To explain 75\% of the variability, $m=8$ principal components are required for defenders, and $m=9$ for midfielders and forwards. These values approximately match those required for all players. Similar to Section~\ref{sec:allgmmcluster}, a diagonal structure was imposed for $\Sigma_k$. For defenders, the BIC was minimised by $K=3$ with cluster sizes of 61, 51 and 30 players. For midfielders, the minimum was at $K=2$. However, the solution allocates defensive players Casemiro, Kalvin Phillips and Yves Bissouma in the same cluster as attacking players such as Marcus Rashford and Joshua Zirkzee, which does not reflect a meaningful tactical distinction. Therefore, the subjective decision was made to interpret the solution under $K=3$ which had a BIC value only 0.41\% greater than for $K=2$. This resulted in cluster sizes of 21, 74 and 49.

For forwards, BIC is minimised at $K = 4$, with cluster sizes of 30, 1, 50 and 31. The singleton cluster consists of Mohamed Salah, who is statistically isolated from all other forwards on the leading principal components (see Section~\ref{sec:forwards_clusters}). His profile lies more than four standard deviations above the positional average on goals plus assists, non-penalty expected goals plus expected assisted goals, and touches in the attacking penalty area. This matches his exceptional individual season, in which he recorded the most goals and assists of any Premier League player, so the singleton cluster reflects an outlier in the data, not an artifact of the model. The remaining three clusters are unchanged from the $K = 3$ solution, which lies 0.53\% above the BIC minimum: at $K = 3$, Salah is absorbed into the creative wide forward cluster, and no other player changes assignment. 

Taken together, the within-position clustering results reveal structural patterns broadly consistent with recognised tactical archetypes, though separation is weaker than in the full-dataset GMM, particularly for midfielders.

\subsubsection{Defender clusters}

Three clusters emerged from the defender dataset. Cluster 1 (61 players; \emph{D: low involvement}) has a comparatively weak and undifferentiated profile with members such as Ben White, Riccardo Calafiori and Tosin Adarabioyo. Its passing completion and aerial duel success sit almost exactly at the positional average, while total touches, live-ball touches, recoveries and passing volume are below average. These defenders are less involved in play and do not strongly express any single stylistic trait.

Cluster 2 (51; \emph{D: wide}) attacking full-backs and wide defenders, such as Trent Alexander-Arnold, Andy Robertson, Jo\v{s}ko Gvardiol and Pedro Porro. It has the highest values on shot-creating actions, key passes, progressive passes received and throw-ins, together with frequent miscontrols, and is lowest on medium and total pass completion, touches in the defensive penalty area, aerial duel success and shots blocked. These players contribute mainly through attacking involvement from wide areas.

Cluster 3 (30; \emph{D: centre backs}) is made up almost entirely of first-choice centre-backs, among them Virgil van Dijk, William Saliba and Ibrahima Konat\'{e}. Touches in the defensive third and defensive penalty area, progressive passing distance and medium-range passing volume are strongly elevated, while attacking and wide actions (shot-creating actions, throw-ins, progressive passes received and crosses) are lowest. They combine penalty area defending with a central role in build-up play. This is the cleanest and most tactically coherent of the three clusters: all players also belong to the centre-back cluster (Cluster 6) of the full-dataset GMM, so the archetype is recovered consistently whether or not the other positions are included.

To illustrate the defensive clusters, Figure~SM1 in the Supplementary Material shows a scatterplot of the first two principal components ($\bu_1$ and $\bu_2$) with plotting symbols coloured by cluster membership.

\subsubsection{Midfielder clusters}

Three clusters emerged from the midfielder dataset. Cluster 1 (21 players; \emph{M: low involvement}) has a shooting-oriented but low-involvement profile. Shots and shots on target per 90 minutes and non-penalty expected goals per shot are positive, while touches, passing volume and live-ball passes are the lowest of any midfielder cluster. It is made up of attacking players classified as midfielders who contribute through shooting with little involvement in build-up play, such as Marcus Rashford, Eddie Nketiah and Joshua Zirkzee.

Cluster 2 (74; \emph{M: creative}), a group of creative and progressive midfielders and attackers with strengths in touches in the attacking third and in shot-creating and goal-creating actions, particularly those arising from live-ball passes. It has no strongly negative features and sits close to or above the positional average on nearly all metrics. The cluster spans a wide range of profiles, from Bukayo Saka and Kevin De Bruyne to Declan Rice and Martin \O{}degaard, which shows the breadth of playing styles that fall under this label. Its size and relatively weak internal differentiation, which points to the stylistic diversity of the midfield role and not a failure of the clustering.

Cluster 3 (49; \emph{M: deep lying}) holds the deeper-lying defensive midfielders, including Casemiro, Kalvin Phillips and Yves Bissouma. They are strongest on shot distance, pass completion, touches in the defensive penalty area and shots blocked, and weakest on attacking output: non-penalty expected goals plus expected assisted goals, goals plus assists, touches in the attacking penalty area and shots on target.

Figure~SM2 in the Supplementary Material shows a scatterplot of the first two principal components ($\bu_1$ and $\bu_2$) with plotting symbols coloured by cluster membership.

\subsubsection{Forward clusters} \label{sec:forwards_clusters}

Four clusters emerged from the forward dataset. Cluster 1 (30 players; \emph{F: low output}) with a comparatively weak profile, made up of lower-output or rotational forwards such as Marcus Rashford, Darwin N\'{u}\~{n}ez and Diogo Jota. None of its features is meaningfully above the positional average, and it records the lowest values on touches, passes received, carries and shot-creating actions, indicating limited involvement in open play.

Cluster 2 (\emph{F: Salah}) is the singleton cluster containing Mohamed Salah. His values lie between 4.2 and 4.7 standard deviations above the positional average on goals plus assists, non-penalty expected goals plus expected assisted goals, touches in the attacking penalty area, assists and goal-creating actions from live-ball passes, and below average on yellow cards, corner kicks taken and shot distance.

Cluster 3 (50; \emph{F: wide}), a group of creative, ball-carrying wide forwards that includes Bukayo Saka, Cole Palmer and Kevin De Bruyne. They have the highest values on total and progressive carrying distance, number of carries and passing volume, and the lowest on non-penalty expected goals per shot and aerial duel involvement. These players contribute mainly through ball progression from wide areas, and their shots are of lower average quality. This is the cluster into which Salah is absorbed at $K = 3$.

Cluster 4 (31; \emph{F: penalty area strikers}) consists of penalty area strikers, for example Erling Haaland, Alexander Isak and Ollie Watkins. It is strongest on aerial duels contested, non-penalty expected goals per shot, offsides and non-penalty expected goals, and weakest on short pass completion, shot distance, shot-creating actions per 90 minutes and crossing. These players operate mainly in the penalty area and shoot from close range, with little involvement in build-up play.

Figure~SM3 in the Supplementary Material shows a scatterplot of the first two principal components ($\bu_1$ and $\bu_2$) with plotting symbols coloured by cluster membership. Mohamed Salah is visible in the top-right hand corner.

\section{Discussion} \label{sec:discussion}

\subsection{Comparison with related work}

\citet{akhanli2023} clustered football players from eight European leagues using a tailor-made dissimilarity measure and a composite validity index framework, finding five positional archetypes consistent with the clusters identified here. The present analysis extends this work in two respects. First, the use of a GMM instead of hard-assignment clustering methods allows hybrid roles to be measured. Second, model selection throughout this paper follows a transparent and repeatable rule: the BIC minimum is adopted wherever it yields interpretable clusters, and any departure from it is justified explicitly and kept within a fraction of a percent of the minimum. This framework could be applied to similar clustering problems in other sports analytics contexts.

\subsection{K-means versus Gaussian mixture models}

K-means is a deterministic algorithm that partitions data by iteratively assigning each observation to the nearest centroid and updating those centroids until convergence. Its principal advantage is computational efficiency and simplicity of interpretation. However, k-means imposes a number of restrictive assumptions: clusters are assumed to be spherical, of roughly equal size, and each observation must belong exclusively to a single group. These constraints mean the algorithm can perform poorly when clusters overlap, vary in shape or density, or when the true structure of the data is inherently ambiguous.

A GMM relaxes these assumptions by modelling the data as arising from a mixture of Gaussian distributions, each with its own mean, covariance, and mixing weight. Rather than assigning observations to clusters deterministically, a GMM produces a probability distribution over cluster membership for each observation, accommodating elliptical cluster shapes and uncertainty at cluster boundaries.

The probabilistic framework offered by the GMM is particularly well suited to the nature of the data. Player performance profiles in the Premier League do not naturally partition into discrete, well-separated groups. Tactical roles exist on a continuum, and many players perform responsibilities that overlap across multiple stylistic archetypes. Applying k-means to this dataset forces every player into a single cluster regardless of how ambiguous their profile might be, which obscures the hybrid-role information this paper sets out to measure. The GMM captures these overlaps explicitly through posterior membership probabilities, as demonstrated in Table~\ref{tab:hybrid_players}.


Fitted to the same $m=8$ principal components with the same number ($K=7$) of clusters, k-means clustering (details not shown) broadly agrees with GMMs (Adjusted Rand Index = 0.495). There is strong one-to-one agreement for Clusters 2 (Defensive midfielders \& full backs), 4 (Set-piece specialists) and 6 (Traditional centre backs). The largest difference lies in how players with less involvement are grouped. K-means keeps defenders and holding midfielders of this type apart from attackers, while the GMM places most of them together in its heterogeneous Cluster 3.

\subsection{Within-position clustering}

Section~\ref{sec:within} demonstrates that GMM is also viable for within-position clustering. This runs contrary to the initial expectation that reduced sample sizes within positional subgroups would favour k-means over GMM, since full-covariance GMM estimation typically requires substantially more observations relative to feature dimensionality to avoid overfitting. By adopting diagonal covariance and a reduced-dimensionality PCA representation retaining 75\% of within-position variance, stable GMM solutions were obtained for all three positions without the overfitting risk associated with full-covariance estimation in smaller samples.

The within-position results also corroborate the full-dataset analysis. All 30 players in the defender centre-back cluster belong to the centre-back cluster of the full-dataset GMM, indicating that this archetype is recovered consistently whether or not other positions are included. The forward analysis identified Mohamed Salah as a singleton cluster at the BIC minimum, lying more than four standard deviations above the positional average on several attacking output measures. This illustrates the ability of the GMM to isolate statistical outliers instead of forcing them into an existing group.

As in the full-dataset analysis, most players within each positional group are assigned to their primary cluster with high confidence. This pattern holds at both stages of the pipeline, which supports using a GMM throughout.

\subsection{Uncertainty and hybrid roles}

The uncertainty analysis identified 16 players (5.1\%) with a maximum posterior probability below 0.65, meaning these players could not be confidently assigned to a single cluster. Some players do not have a clear positional identity in statistical terms, and the posterior probabilities provide a quantitative measure of this ambiguity. The two measures of ambiguity used here agree closely, with 15 of these 16 players also meeting the 25\% hybrid membership threshold. Mikel Merino is the clearest example, with his membership divided almost equally across three clusters. A similar, smaller pattern was observed within each positional group, with between 0.9\% and 4.2\% of players falling below the 0.65 threshold depending on position.

Assignment uncertainty therefore persists within positions, although at a lower rate. Stylistic ambiguity in modern football appears to operate at multiple levels: broadly, in terms of which archetype a player most resembles, and more finely, in terms of which sub-archetype within a position best describes their playing style.

Statistical hybridity is largely distinct from positional ambiguity in the source data. Of the 84 players listed in more than one positional group, only 8 meet the 25\% hybrid threshold, a rate barely above that of single-position players, while many of the strongest hybrids, such as Cristian Romero, Malo Gusto and Dominik Szoboszlai, are listed in a single position. A player's listed position describes where they are deployed; GMM hybridity describes how their statistical profile relates to the archetypes identified, and the two should not be treated as interchangeable.

\subsection{Limitations} \label{sec:limitations}

The dataset covers a single Premier League season, which means the results reflect one snapshot of player performance and may not generalise across seasons, leagues, or tactical contexts. For example, a player recovering from injury or adapting to a new system mid-season will have a statistical profile that blends contributions from different contexts, introducing noise that the model cannot distinguish from genuine stylistic ambiguity.

With the exception of rate and percentage variables, the features used for clustering are season totals, and players were required to have played at least 800 minutes. Because counting statistics scale with minutes played, the clustering captures the volume of a player's involvement across the season as well as their playing style. This is visible in the results: in each positional group, one cluster consists predominantly of players with substantially fewer minutes than the remainder (median 1,198 versus 2,614 minutes for defenders, 1,034 versus 1,965 for midfielders and 994 versus 2,200 for forwards; one-sided Mann--Whitney U tests, all $p < 10^{-10}$), and the remaining clusters are also ordered by playing time. As a sensitivity analysis, all counting variables were converted to per-90-minute rates and the models were refitted. Under this specification BIC favoured $K = 2$ in every positional group, yielding a coarse division that did not resolve the tactical archetypes identified in the main analysis. The season-total specification was therefore retained, but the resulting clusters should be interpreted as profiles of both playing style and season-long involvement and not style alone.

The feature set, while broad, is limited to what FBref makes publicly available. Metrics such as pressing intensity, off-ball movement, defensive positioning, and spatial coverage are increasingly central to modern tactical analysis but are not captured in publicly available aggregated statistics. Their absence likely flattens authentic stylistic differences between players, particularly within the midfielder cluster where separation was weakest.

The cluster labels assigned throughout were applied after clustering and are inherently subjective. Two analysts examining the same cluster profiles might reach different labelling decisions, and there is no ground truth against which to evaluate them. This is a fundamental limitation of unsupervised learning in interpretive contexts as opposed to a flaw specific to this analysis.

\subsection{Extensions} \label{sec:extensions}

A natural extension for this work would be building a player similarity tool on top of the GMM posterior probabilities. Rather than asking which cluster a player belongs to, one could compute the distance between any two players' full posterior distributions and identify the statistically most similar player to any given individual. Finding the closest statistical match to Mohamed Salah, or identifying which players in the league most resemble a recently departed transfer target, would give the framework direct practical recruitment value.

Another extension would be tracking cluster membership across multiple seasons to map player development quantitatively. A midfielder whose posterior probability shifts gradually from a defensive cluster toward a creative one over three seasons is showing a statistically measurable change in role, which could be useful for both development analysis and recruitment planning.

From a modelling perspective, a Dirichlet process mixture model would remove the need to specify the number of components in advance, allowing the data to determine the appropriate number of clusters directly and addressing one of the more arbitrary decisions in the current pipeline.

\section*{Supplementary Material}

The supplementary material can be found at \newline \url{https://www.personal.soton.ac.uk/amo105/clusteringPL_supplementary.html}.

\bibliographystyle{plainnat}
\bibliography{references}

\end{document}